\documentclass[a4paper,11pt]{article}

\usepackage{authblk}
\usepackage{algorithm}
\usepackage{algorithmic}
\usepackage{multirow}
\usepackage{graphicx}
\usepackage{booktabs}
\usepackage{balance}
\usepackage{epsfig}
\usepackage{epstopdf}
\usepackage{amsmath}
\usepackage{amsthm}
\usepackage{amsfonts}
\usepackage{subcaption}
\usepackage{wasysym}
\usepackage{microtype}
\usepackage{rotating}
\usepackage{tikz}
\usepackage{txfonts}
\usepackage{fancyhdr}
\usepackage{url}
\usepackage{hyperref}
\usetikzlibrary{positioning}
\usetikzlibrary{decorations.pathreplacing}
\tikzset{
	mybrace/.style={decorate,decoration={brace,aspect=#1}}
}

\newcommand{\N}{\mathbb{N}}

\newcommand{\F}{\mathbb{F}}

\newtheorem{definition}{Definition}
\newtheorem{lemma}{Lemma}

\newtheorem{theorem}{Theorem}

\newtheorem{remark}{Remark}
\newtheorem{problem}{Problem}
\newtheorem{example}{Example}
\newtheorem{corollary}{Corollary}

\providecommand{\keywords}[1]{\textbf{\textit{Keywords }} #1}

\begin{document}

\title{Latin Hypercubes and Cellular Automata}

\author[1]{Maximilien Gadouleau}
\author[2]{Luca Mariot}

\affil[1]{{\normalsize Department of Computer Science, Durham University, South
              Road, Durham DH1 3LE, United Kingdom} \\
  
    {\small \texttt{m.r.gadouleau@durham.ac.uk}}}
\affil[2]{{\normalsize Cyber Security Research Group, Delft University of Technology,
              Mekelweg 2, Delft, The Netherlands} \\
  
    {\small \texttt{l.mariot@tudelft.nl}}}

\maketitle

\begin{abstract}
  Latin squares and hypercubes are combinatorial designs with several applications in statistics, cryptography and coding theory. In this paper, we generalize a construction of Latin squares based on bipermutive cellular automata (CA) to the case of Latin hypercubes of dimension $k>2$. In particular, we prove that linear bipermutive CA (LBCA) yielding Latin hypercubes of dimension $k>2$ are defined by sequences of invertible Toeplitz matrices with partially overlapping coefficients, which can be described by a specific kind of regular de Bruijn graph induced by the support of the determinant function. Further, we derive the number of $k$-dimensional Latin hypercubes generated by LBCA by counting the number of paths of length $k-3$ on this de Bruijn graph.
\end{abstract}


\section{Introduction}
\label{sec:intro}
Several cryptographic protocols with information-theoretic security guarantees can be defined in terms of \emph{Combinatorial Designs}. One such example are $(k,n)$ threshold \emph{Secret Sharing Schemes} (SSS), where a \emph{dealer} wants to share a secret information $S$ among a set of $n$ \emph{players} by giving to each of them a \emph{share} $B$, so that at least $k$ players are required to recover $S$ by combining their shares. On the other hand, any subset of $k-1$ or less players do not gain any information on the secret by pooling the respective shares. Such a protocol corresponds to a set of $n$ \emph{Mutually Orthogonal Latin Squares} (MOLS) when $k=2$ players are required to reconstruct the secret, while it is equivalent to a set of $n$ \emph{Mutually Orthogonal Latin Hypercubes} of dimension $k$ (MOLH) when $k>2$~\cite{stinson04}. Indeed, the schemes proposed by Shamir~\cite{shamir79} and Blakley's~\cite{blakley79} can be thought as particular instances of $(k,n)$ threshold SSS, where polynomials and hyperplanes over finite fields are respectively used to represent an underlying set of MOLH in a compact way.

A recent research thread considers the design of secret sharing schemes by means of \emph{Cellular Automata} (CA)~\cite{mariot14}, the goal being twofold. First, a CA-based architecture could be useful for efficient hardware-oriented implementations of threshold SSS, by leveraging the massive parallelism of the CA models. This approach already turned out to be interesting in another cryptographic application, namely the design of \emph{S-boxes} based on CA~\cite{mariot19}. Second, the locality of the CA model can be used to define access structures that are more constrained than the classic $(k,n)$ threshold, such as the \emph{consecutive access structure} of the CA-based scheme proposed in~\cite{mariot14}, which finds application in certain distributed cryptographic protocols, as discussed by the authors of~\cite{herranz18}. These access structures eventually become \emph{cyclic}, and the maximum number of players allowed is related to the period of spatially periodic preimages in surjective CA~\cite{mariot17}.

Given the equivalence between $(k,n)$ threshold SSS and families of MOLH, a possible way to tackle these goals is to study how to construct the latter using cellular automata. Recently, the authors of~\cite{mariot20} showed a construction of MOLS based on \emph{linear bipermutive} CA (LBCA), thereby addressing the case of $(2, n)$ threshold SSS based on CA. Generalizing this construction to a higher threshold $k$ entails two steps: first, one needs to characterize which CA generate Latin hypercubes, since contrarily to the Latin squares case not all LBCA define Latin hypercubes of dimension $k>2$. The next step is to define subsets of such CA whose Latin hypercubes are $k$-wise orthogonal, thus obtaining sets of $MOLH$.

The aim of this paper is to address the first step of this generalization, namely the characterization and enumeration of Latin hypercubes based on LBCA. In particular, we first prove that LBCA which generate \emph{Latin cubes} are defined by local rules whose central coefficients compose the border of an invertible \emph{Toeplitz matrix}. This allows us to determine the number of Latin cubes generated by LBCA by counting the number of invertible Toeplitz matrices. Next, we generalize this result to dimension $k>3$, remarking that in this case the local rule of an LBCA can be defined by a \emph{sequence} of $k-2$ invertible Toeplitz matrices, where adjacent matrices share the coefficients respectively of the upper and lower triangulars. We finally show that this overlapping relation can be described by a regular de Bruijn graph, and that paths of length $k-2$ on this graph corresponds to LBCA generating $k$-dimensional Latin hypercubes.

The rest of this paper is organized as follows. Section~\ref{sec:prelim} covers the necessary background on Latin hypercubes and cellular automata used in the paper, and recalls the basic results about Latin squares generated by CA. Section~\ref{sec:char} presents the characterization of Latin cubes and hypercubes generated by LBCA defined by invertible Toeplitz matrices. Section~\ref{sec:count} proves the regularity of the de Bruijn graph associated to invertible Toeplitz matrices and derives the number of $k$-dimensional Latin hypercubes generated by LBCA. Section~\ref{sec:conc} recaps the main contributions of the paper and discusses some directions for future research.

\section{Preliminary Definitions and Results}
\label{sec:prelim}
The main combinatorial object of interest of this paper are \emph{Latin hypercubes}, which we define as follows:

\begin{definition}
	\label{def:lat-hc}
	Let $X$ be a finite set of $N \in \N$ elements. A \emph{Latin hypercube} of dimension $k \in \N$ and order $N$ over $X$ is a $k$-dimensional array with entries from $X$ such that, by fixing any subset $i_1, \cdots, i_{k-1}$ of $k-1$ coordinates, the remaining coordinate $i_k$ yields a permutation over $X$, that is, each element in $X$ appears exactly once on the $i_k$ coordinate.
\end{definition}
Remark that when $k=2$ one obtains the definition of \emph{Latin square} of order $N$, i.e. a square matrix where each row and each column is a permutation of $X$. Usually, Latin squares and hypercubes are defined over the set $X = [N] = \{1,\cdots, N\}$ of the first $N$ positive integers. In this work, we consider the case where $X$ is the vector space $\F_q^b$, with $\F_q$ being the finite field of $q$ elements. Hence, the order $N$ of the hypercube will be $q^b$, i.e. the number of all $q$-ary vectors of length $b$.

Next, we define the model of \emph{cellular automaton} used in the rest of this paper:
\begin{definition}
	\label{def:nbca}
	Let $d, n \in \N$, with $d\le n$, and let $f: \F_q^d \rightarrow \F_q$ be a function of $d$ variables over the finite field $\F_q$, with $q$ being a power of a prime. The \emph{Cellular Automaton} (CA in the following) of length $n$ and local rule $f$ over the alphabet $\F_q$ is the vectorial function $F: \F_q^n \rightarrow \F_q^{n-d+1}$ defined for all vectors $x = (x_1,\cdots,x_n) \in \F_q^n$ as:
	\begin{equation}
	\label{eq:glob-rule}
	F(x_1,\cdots,x_n) = (f(x_1,\cdots,x_d), \cdots, f(x_{n-d+1},\cdots, x_n)) \enspace .
	\end{equation}
\end{definition}
In other words, each coordinate function $f_i:\F_q^n \rightarrow \F_q$ for $i \in [n-d+1]$ of a CA $F$ corresponds to its local rule $f$ applied to the neighborhood $\{x_i,\cdots,x_{i+d-1}\}$. For our examples, in the following we will mainly consider the case $q=2$, where a CA boils down to a particular type of \emph{vectorial Boolean function} $F: \F_2^n \rightarrow \F_2^{n-d+1}$. This corresponds to the definition of \emph{no-boundary cellular automaton} (NBCA) studied in~\cite{mariot19} as a model for cryptographic S-boxes.

\begin{figure}[t]
	\centering
	\begin{tikzpicture}[every node/.style={font=\sffamily\bfseries,minimum size=0.5cm},
	rect node/.style={rectangle,draw,font=\sffamily\bfseries,minimum size=0.5cm, inner sep=0pt, outer sep=0pt},
	rect1 node/.style={rectangle,draw,fill=gray!70,font=\sffamily\bfseries,minimum size=0.5cm, inner sep=0pt, outer sep=0pt},
	empt node/.style={font=\sffamily,inner sep=0pt,minimum size=0pt}, on grid]
	
	\node [rotate=45] at (0.45,2.15) {{\scriptsize $y\!=\!1$}};
	\node [rotate=45] at (0.95,1.8) {{\scriptsize $y\!=\!2$}};
	\node [rotate=45] at (1.45,1.45) {{\scriptsize $y\!=\!3$}};
	\node [rotate=45] at (1.9,1.1) {{\scriptsize $y\!=\!4$}};
	
	\node at (-0.4,1.75) {{\scriptsize $x\!=\!1$}};
	\node at (-0.4,1.25) {{\scriptsize $x\!=\!2$}};
	\node at (-0.4,0.75) {{\scriptsize $x\!=\!3$}};
	\node at (-0.4,0.25) {{\scriptsize $x\!=\!4$}};
	
	\node at (2,-1.6) {{\scriptsize $z\!=\!1$}};
	\node at (4.25,-1.65) {{\scriptsize $z\!=\!2$}};
	\node at (6.5,-1.7) {{\scriptsize $z\!=\!3$}};
	\node at (8.5,-1.8) {{\scriptsize $z\!=\!4$}};
	
	\begin{scope}[every node/.append style={yslant=-0.7},yslant=-0.7]
	\node at (0.25,1.75) {1};
	\node at (0.75,1.75) {2};
	\node at (1.25,1.75) {3};
	\node at (1.75,1.75) {4};
	\node at (0.25,1.25) {2};
	\node at (0.75,1.25) {1};
	\node at (1.25,1.25) {4};
	\node at (1.75,1.25) {3};
	\node at (0.25,0.75) {3};
	\node at (0.75,0.75) {4};
	\node at (1.25,0.75) {1};
	\node at (1.75,0.75) {2};
	\node at (0.25,0.25) {4};
	\node at (0.75,0.25) {3};
	\node at (1.25,0.25) {2};
	\node at (1.75,0.25) {1};
	\draw [step=0.5] (0,0) grid (2,2);
	\end{scope}
	\begin{scope}[every node/.append style={yslant=-0.7},yslant=-0.7]
	\node at (2.5,3.25) {2};
	\node at (3,3.25) {1};
	\node at (3.5,3.25) {4};
	\node at (4,3.25) {3};
	\node at (2.5,2.75) {1};
	\node at (3,2.75) {2};
	\node at (3.5,2.75) {3};
	\node at (4,2.75) {4};
	\node at (2.5,2.25) {4};
	\node at (3,2.25) {3};
	\node at (3.5,2.25) {2};
	\node at (4,2.25) {1};
	\node at (2.5,1.75) {3};
	\node at (3,1.75) {4};
	\node at (3.5,1.75) {1};
	\node at (4,1.75) {2};
	\draw [step=0.5,xshift=2.25cm, yshift=1.5cm](0,0) grid (2,2);
	\end{scope}
	\begin{scope}[every node/.append style={yslant=-0.7},yslant=-0.7]
	\node at (4.75,4.75) {3};
	\node at (5.25,4.75) {4};
	\node at (5.75,4.75) {1};
	\node at (6.25,4.75) {2};
	\node at (4.75,4.25) {4};
	\node at (5.25,4.25) {3};
	\node at (5.75,4.25) {2};
	\node at (6.25,4.25) {1};
	\node at (4.75,3.75) {1};
	\node at (5.25,3.75) {2};
	\node at (5.75,3.75) {3};
	\node at (6.25,3.75) {4};
	\node at (4.75,3.25) {2};
	\node at (5.25,3.25) {1};
	\node at (5.75,3.25) {4};
	\node at (6.25,3.25) {3};
	\draw [step=0.5,xshift=4.5cm, yshift=3cm](0,0) grid (2,2);
	
	\end{scope}
	\begin{scope}[every node/.append style={yslant=-0.7},yslant=-0.7]
	\node at (7,6.25) {4};
	\node at (7.5,6.25) {3};
	\node at (8,6.25) {2};
	\node at (8.5,6.25) {1};
	\node at (7,5.75) {3};
	\node at (7.5,5.75) {4};
	\node at (8,5.75) {1};
	\node at (8.5,5.75) {2};
	\node at (7,5.25) {2};
	\node at (7.5,5.25) {1};
	\node at (8,5.25) {4};
	\node at (8.5,5.25) {3};
	\node at (7,4.75) {1};
	\node at (7.5,4.75) {2};
	\node at (8,4.75) {3};
	\node at (8.5,4.75) {4};
	\draw [step=0.5,xshift=6.75cm, yshift=4.5cm](0,0) grid (2,2);
	\end{scope}
	
	\node[empt node] (e1) {};
	\node[empt node] (e2) [right=4cm of e1] {};
	\node[empt node] (e3) [above=2.6cm of e2] {};
	\node[empt node] (e4) [above=1.1cm of e2] {};
	\node[empt node] (e5) [left=0.45cm of e4] {};
	\node [rect node] (c1) [above=3.5cm of e2] {1};
	\node [rect node] (c2) [right=0cm of c1.east] {1};
	
	\node [empt node] (f1) [above=0.35cm of c1.east] {$\Downarrow$};
	\node [empt node] (f3) [right=0.25cm of f1] {$F$};
	
	\node [rect node] (p2) [above=1cm of c1] {0};
	\node [rect node] (p1) [left=0cm of p2.west] {0};
	\node [rect node] (p1a) [left=0cm of p1.west] {0};
	\node [rect node] (p3) [right=0cm of p2.east] {1};
	\node [rect node] (p4) [right=0cm of p3.east] {1};
	\node [rect node] (p5) [right=0cm of p4.east] {0};
	
	\draw [-, decorate, decoration={brace,mirror,amplitude=5pt,raise=0.3cm}]
	(c1.west) -- (c2.east) node [midway,yshift=-0.7cm] {$F(1,3,2)=4$};
	
	\draw [-, decorate, decoration={brace,amplitude=5pt,raise=0.3cm}]
	(p1a.west) -- (p1.east) node [midway,yshift=0.7cm] {$x=1$};
	\draw [-, decorate, decoration={brace,amplitude=5pt,raise=0.3cm}]
	(p2.west) -- (p3.east) node [midway,yshift=0.7cm] {$y=3$};
	\draw [-, decorate, decoration={brace,amplitude=5pt,raise=0.3cm}]
	(p4.west) -- (p5.east) node [midway,yshift=0.7cm] {$z=2$};
	
	\draw[->,thick, shorten >=0pt,shorten <=0pt,>=stealth] (e3) -- (e5);
	
	\end{tikzpicture}
	\caption{Latin cube of order $4$ generated by a CA $F:\F_2^6 \rightarrow \F_2^2$ defined by local rule $f(x_1,\cdots, x_5) = x_1 \oplus x_3 \oplus x_5$. The encoding used is $00 \mapsto 1$, $10 \mapsto 2$, $01 \mapsto 3$, $11 \mapsto 4$.}
	\label{fig:ex-lat-cube}
\end{figure}
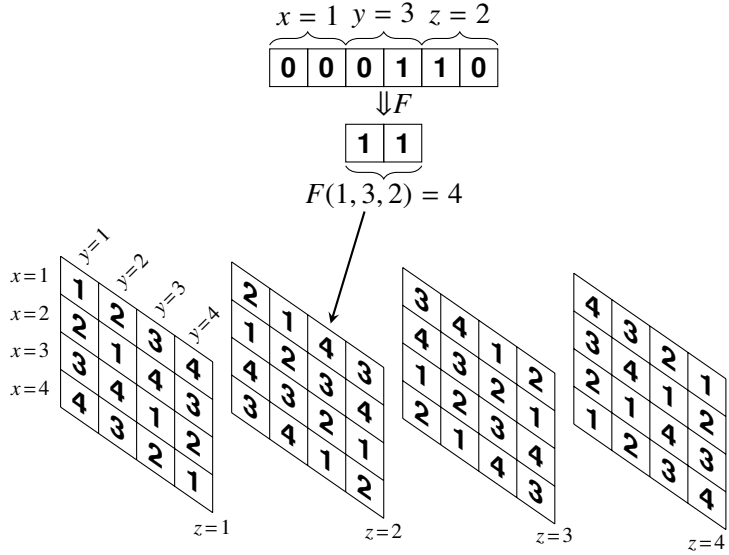

The last preliminary definition we need is that of hypercube associated to a CA. In what follows, we assume that the vectors in $\F_q^b$ are totally ordered, and that there is a monotone and one-to-one mapping $\Psi: [N]\rightarrow \F_q^b$, in order to associate sets of integer coordinates to sets of $q$-ary vectors.
\begin{definition}
	\label{def:hc-ca}
	Let $b, k \in \N$, with $d = b(k-1)+1$. Moreover, let $F: \F_q^{bk} \rightarrow \F_q^b$ be the CA of length $bk$ defined by a local rule $f:\F_q^d \rightarrow \F_q$. Then, the \emph{hypercube associated to} $F$ of order $N = q^b$ is the $k$-dimensional array $\mathcal{H}_F$ where for all vectors of coordinates $(i_1,\cdots,i_k) \in [N]^k$ the corresponding entry is defined as:
	\begin{equation}
	\label{eq:hc-ca}
	\mathcal{H}_F(i_1,\cdots,i_k) = \Psi^{-1}(F(\Psi(i_1) || \Psi(i_2) || \cdots || \Psi(i_k)) \enspace ,
	\end{equation}
	where the input $\Psi(i_1) || \Psi(i_2) || \cdots || \Psi(i_k)$ denotes the \emph{concatenation} of the binary vectors $\Psi(i_1), \cdots, \Psi(i_k) \in \F_q^b$.
\end{definition}
Thus, by Definition~\ref{def:hc-ca} the hypercube associated to a CA $F$ of length $bk$ and local rule $f$ of $b(k-1)+1$ is constructed by splitting the input vector of $F$ in $k$ blocks of size $b$, which are used to index the coordinates of the hypercube, while the output vector represents the entry to be put at those coordinates. Figure~\ref{fig:ex-lat-cube} depicts an example of $3$-dimensional hypercube $\mathcal{H}_F$ of block size $b=2$ over $\F_2$ (thus, of order $2^2=4$) associated to the CA $F:\F_2^6 \rightarrow \F_2^2$ which is defined by the local rule $f(x_1,\cdots, x_5) = x_1 \oplus x_3 \oplus x_5$. In this case one can see that $\mathcal{H}_F$ is indeed a Latin \emph{cube}, i.e. a Latin hypercube of dimension 3.

We now give the formal statement of the problem investigated in this paper, namely the construction and enumeration of Latin hypercubes with CA:
\begin{problem}
	\label{prob:stat}
	Let $F: \F_q^{bk} \rightarrow \F_q^b$ be a CA equipped with a local rule $f:\F_2^d \rightarrow \F_2$ where $d = b(k-1)+1$. When is the hypercube $\mathcal{H}_F$ associated to $F$ a Latin hypercube? How many local rules $f: \F_q^{b(k-1)+1} \rightarrow \F_q$ induce a CA such that the resulting $\mathcal{H}_F$ is a Latin hypercube? 
\end{problem}
Problem~\ref{prob:stat} requires studying under which conditions the local rule $f$ induces a permutation between any of the $k$ blocks of $b$ consecutive cells used to index the coordinates of $\mathcal{H}_F$ and the output CA configuration, when all remaining blocks are fixed. We first start by addressing the extremal cases of the leftmost and rightmost blocks in CA defined by \emph{bipermutive} local rules.

A function $f: \F_q^n \rightarrow \F_q$ of $n\ge 2$ variables is \emph{bipermutive} if there exists a function $g: \F_q^{n-2}\rightarrow \F_q$ such that
\begin{equation}
\label{eq:bip-func}
f(x_1, \cdots , x_n) = x_1 \oplus g(x_2, \cdots , x_{n-1}) \oplus
x_n
\end{equation}
for all $x = (x_1,\cdots,x_n) \in \F_q^n$, where $\oplus$ corresponds to the sum operation over $\F_q$. When $q=2$, Equation~\eqref{eq:bip-func} basically amounts to the XOR of the leftmost and rightmost input variables with the result of function $g$ computed on the central $n-2$ variables. A proof of the following result can be found in~\cite{mariot20}:
\begin{lemma}
	\label{lm:hc-left-right}
	Let $F: \F_q^{bk} \rightarrow \F_q^b$ be a CA with bipermutive rule $f:\F_q^d \rightarrow \F_q$, where $d = b(k-1)+1$. Then, the restriction $F|_{\tilde{x}}:\F_q^b \rightarrow \F_q^b$ obtained by fixing either the rightmost or leftmost $b(k-1)$ variables in the CA input to $\tilde{x} \in \F_q^{b(k-1)}$ is a permutation over $\F_q^{b}$ for all $\tilde{x} \in \F_q^{b(k-1)}$.
\end{lemma}
Thus, by Lemma~\ref{lm:hc-left-right} bipermutivity of the local rule $f$ is a sufficient condition for verifying the Latin hypercube property on the leftmost and rightmost coordinate of the hypercube $\mathcal{H}_F$. This also means, in turn, that for dimension $k=2$ (that is, when there are no blocks in the middle between the leftmost and the rightmost one) bipermutivity is a sufficient condition to ensure that $\mathcal{H}_F$ is a Latin square:
\begin{corollary}
	\label{cor:bip-lat-sq}
	Let $F: \F_q^{2b} \rightarrow \F_q^b$ be a CA with bipermutive rule $f: \F_q^d \rightarrow \F_q$ of diameter $d = b+1$. Then, the hypercube $\mathcal{H}_F$ associated to $F$ is a Latin square of order $q^b$.
\end{corollary}
From Corollary~\ref{cor:bip-lat-sq}, one also gets the following straightforward counting result:
\begin{corollary}
	\label{cor:count-lsq}
	Let $b \in \N$. Then, the number of Latin squares of order $q^b$ generated by bipermutive CA corresponds to the number of bipermutive local rules of $b+1$ variables over $\F_q$, which is $q^{q^{b-1}}$.
\end{corollary}
Hence, Corollaries~\ref{cor:bip-lat-sq} and~\ref{cor:count-lsq} solve Problem~\ref{prob:stat} for dimension $k=2$. In what follows, we shall solve Problem~\ref{prob:stat} in Theorems~\ref{thm:lat-hc-ca} and~\ref{thm:num-hc-ca} for any dimension $k>2$.

\section{CA-based Latin Hypercubes from Toeplitz Matrices}
\label{sec:char}

\subsection{Latin Cubes}
\label{subsec:cubes}
Remark that for dimension $k>2$ bipermutivity is not enough. As a matter of fact, Lemma~\ref{lm:hc-left-right} requires the $b(k-1)$ variables to be adjacent. To verify the Latin hypercube property for a middle coordinate $1<i<k$ one needs to fix all variables on the left and on the right except for the ``hole'' represented by the block of $b$ bits associated to coordinate $i$. As a consequence, it is necessary to characterize a proper subset of bipermutive local rules that generate Latin hypercubes when used as local rules of CA.

We begin by addressing the case of \emph{Latin cubes}, that is with dimension $k=3$. Referring to Problem~\ref{prob:stat}, we have a CA $F:\F_q^{3b} \rightarrow \F_q^b$ that maps configurations of $3b$ cells in vectors of $b$ cells, defined by a bipermutive local rule $f: \F_q^{d}\rightarrow \F_q$ of diameter $d=2b+1$. Since the permutation between the output CA configuration and the blocks $x_{[1,b]}$ and $x_{[2b+1,3b]}$ is already granted by Lemma~\ref{lm:hc-left-right}, we only need to consider the middle block $x_{[b+1,2b]}$.

In what follows, we will also make the additional assumption that, beside being bipermutive, the local rule is \emph{linear}: In other words, there exist a binary vector $a = (a_1,a_2,\cdots, a_{d-1},a_d) \in \F_q^{d}$ such that
\begin{equation}
\label{eq:lin-bip}
f(x_1,x_2, \cdots, x_{d-1},x_d) = a_1x_1 \oplus a_2x_2 \oplus \cdots  \oplus a_{d-1}x_{d-1} \oplus a_dx_d \enspace ,
\end{equation}
where sum and product are considered over $\F_q$. Notice that a linear rule defined as in~\eqref{eq:lin-bip} is bipermutive if and only if both $a_1$ and $a_d$ are not null. In particular, from now on we will assume that $a_1=a_d=1$, and we will define a linear bipermutive rule by means of the vector $(a_2,\cdots,a_{d-1}) \in \F_q^{d-2}$ defining the $d-2$ central coefficients. Additionally, we will refer to a CA defined by such a rule as a LBCA (Linear Bipermutive CA).

For all $x \in \F_q^{3b}$, let $y = F(x) \in \F_q^b$ be the result of the CA applied to vector $x$. Since the local rule $f$ is linear, one can express $y$ as a system of $b$ linear equations and $3b$ variables:
\begin{equation}
\label{eq:sys-lin-bip}
\begin{cases}
y_1 &= x_1 \oplus a_2x_2 \oplus \cdots \oplus a_{2b}x_{2b} \oplus x_{2b+1} \\
y_2 &= x_2 \oplus a_2x_3 \oplus \cdots \oplus a_{2b}x_{2b+1} \oplus x_{2b+2} \\
&\vdots \\
y_b &= x_{b} \oplus a_2x_{b+1} \oplus \cdots \oplus a_{2b}x_{3b-1} \oplus x_{3b}
\end{cases}
\end{equation} 
Suppose now that we fix the $2b$ variables $x_1,\cdots, x_b$ and $x_{2b+1},\cdots,x_{3b}$ respectively to the values $\tilde{x}_1,\cdots, \tilde{x}_b$ and $\tilde{x}_{2b+1},\cdots, \tilde{x}_{3b}$. This actually amounts to fixing the leftmost and the rightmost coordinates in the cube $\mathcal{H}_F$ associated to $F$. Moreover, the system~\eqref{eq:sys-lin-bip} becomes a system of $b$ linear equations and $b$ variables corresponding to the block $x_{[b+1,2b]}$, since the remaining $2b$ variables have been set to constant values. In order to ensure that there is a permutation between $x_{[b+1,2b]}$ and $y$, it means that the matrix of coefficients $a_i$ multiplying the vector $x_{[b+1,2b]}$ in~\eqref{eq:sys-lin-bip} must be invertible:
\begin{equation}
\label{eq:toeplitz}
M_F =
\begin{pmatrix}
a_{b+1} & a_{b+2} & \cdots & a_{2b} \\
a_{b} & a_{b+1} & \cdots & a_{2b-1} \\
\vdots  & \vdots & \ddots & \vdots \\
a_2 & a_3 & \cdots & a_{b+1}
\end{pmatrix}
\end{equation}
Remark that the matrix in Equation~\eqref{eq:toeplitz} is a \emph{Toeplitz matrix}, where the first row of coefficients $a_{b+1}, \cdots, a_{2b}$ is shifted to the right while the coefficients $a_{b},\cdots, a_2$ progressively enter from the left. In particular, the matrix is completely characterized by the shifts of the central coefficients $a_2,\cdots, a_b, \cdots, a_{2b}$ of the CA local rule $f$. To summarize, we obtained the following result:
\begin{lemma}
	\label{lm:lat-cube}
	Let $F:\F_q^{3b} \rightarrow \F_q^b$ be a LBCA with rule $f: \F_q^{2b+1}\rightarrow \F_q$ defined for
	all $x \in \F_q^{2b+1}$ as
	\begin{displaymath}
	f(x_1,\cdots,x_{2b+1}) = x_1 \oplus a_2x_2 \oplus \cdots \oplus a_{2b}x_{2b} \oplus x_{2b+1} \enspace .
	\end{displaymath}
	Then, the hypercube $\mathcal{H}_F$ associated to $F$ is a Latin cube of order $q^b$ if and only if the Toeplitz matrix $M_F$ defined by the coefficients $a_2,\cdots, a_{2b} \in \F_q$ is invertible.
\end{lemma}
The authors of~\cite{armas11} showed that the number of nonsingular $n\times n$ Toeplitz matrices is $q^{2(n-1)}(q-1)$. Hence, we have the following counting result for Latin cubes:
\begin{theorem}
	\label{thm:num-lat-cub}
	Let $b \in \N$. Then, the number of LBCA $F:\F_q^{3b}\rightarrow \F_q^b$ whose associated hypercube $\mathcal{H}_F$ is a Latin cube is $q^{2(b-1)}(q-1)$.
\end{theorem}

\subsection{Latin Hypercubes of Dimension $k>3$}
\label{subsec:hyper}
We now generalize the investigation to Latin hypercubes of any dimension $k>3$. In this case, the LBCA $F:\F_q^{bk} \rightarrow \F_q^{b}$ is defined by a rule $f:\F_q^{b(k-1)+1} \rightarrow \F_q$ of the form:
\begin{equation}
\label{eq:lin-rule-hc}
f(x_1,\cdots, x_{b(k-1)+1}) = x_1 \oplus a_2x_2 \oplus \cdots \oplus a_{b(k-1)}x_{b(k-1)} \oplus x_{b(k-1)+1} \enspace .
\end{equation}
Hence, the values of the output vector $y = F(x) \in \F_q^b$ will be determined by a system analogous to~\eqref{eq:sys-lin-bip}:
\begin{equation}
\label{eq:sys-lin-bip-2}
\begin{cases}
y_1 &= x_1 \oplus a_2x_2 \oplus \cdots \oplus a_{b(k-1)}x_{b(k-1)} \oplus x_{b(k-1)+1} \\
y_2 &= x_2 \oplus a_2x_3 \oplus \cdots \oplus a_{b(k-1)}x_{b(k-1)+1} \oplus x_{b(k-1)+2} \\
&\vdots \\
y_b &= x_{b} \oplus a_2x_{b+1} \oplus \cdots \oplus a_{b(k-1)}x_{bk-1} \oplus x_{bk}
\end{cases}
\end{equation}
The $k$-dimensional hypercube $\mathcal{H}_F$ associated to $F$ will be a Latin hypercube only if there is a permutation between any of the central $k-2$ blocks of $b$ cells when all the others are fixed to a constant value and $y$ (the leftmost
and rightmost cases already being granted by bipermutivity). Similarly to the three-dimensional case where we had only one central block, this means that all of the following Toeplitz matrices must be invertible for all $i \in [k-2] = \{1,\cdots, k-2\}$:
\begin{equation}
\label{eq:toep-mat-gen} 
M_{F,i} =
\begin{pmatrix}
a_{bi+1} & a_{bi+2} & \cdots & a_{b(i+1)-1} \\
a_{bi} & a_{bi+1} & \cdots & a_{b(i+1)-2} \\
\vdots  & \vdots & \ddots & \vdots \\
a_{b(i-1)+2} & a_{b(i-1)+3} & \cdots & a_{bi+1}
\end{pmatrix}
\end{equation}
where $M_{F,1}$ is associated to the permutation on the second block, $M_{F,2}$ to the permutation on the third, and so on until $M_{F,k-2}$, which is associated to the permutation on the $(k-1)$-th block. We thus get the following characterization of LBCA that generate $k$-dimensional Latin hypercubes:
\begin{theorem}
	\label{thm:lat-hc-ca}
	Let $F: \F_2^{bk}\rightarrow \F_2^b$ be a CA with local rule $f:\F_2^{b(k-1)+1} \rightarrow \F_2$ defined as in
	Equation~\eqref{eq:lin-rule-hc}. Then, the $k$-dimensional hypercube $\mathcal{H}_F$ of order $q^b$ associated to $F$ is a Latin hypercube if and only if the Toeplitz matrix $M_{F,i}$ in~\eqref{eq:toep-mat-gen} is invertible for all $i \in [k-2]$.
\end{theorem}

Since we settled the first part of Problem~\ref{prob:stat}, we now focus on the counting question, i.e. what is the number $L_{b,k}$ of LBCA $F:\F_q^{bk} \rightarrow \F_q^b$ that generate $k$-dimensional Latin hypercubes of order $q^b$. In other words, we want to count in how many ways the coefficients $a_2,\cdots,a_{b(k-1)} \in \F_2$ of the local rule $f$ in~\eqref{eq:lin-rule-hc} can be chosen so that the resulting hypercube $\mathcal{H}_F$ is a Latin hypercube. 

\begin{remark}
	\label{rem:overlap}
	For dimensions $k=2$ and $k=3$, we settled the counting question of Problem~\ref{prob:stat} respectively in Corollaries~\ref{cor:count-lsq} and~\ref{thm:num-lat-cub}. For dimension $k>3$, remark that one cannot choose independently the coefficients defining the $k-2$ invertible Toeplitz matrices. Indeed, consider the two adjacent matrices $M_{F,i}$ and $M_{F,i+1}$:
	\begin{displaymath}
	\begin{pmatrix}
	a_{bi+1} & a_{bi+2} & \cdots & a_{b(i+1)} \\
	a_{bi} & a_{bi+1} & \cdots & a_{b(i+1)-1} \\
	\vdots  & \vdots & \ddots & \vdots \\
	a_{b(i-1)+2} & a_{b(i-1)+3} & \cdots & a_{bi+1}
	\end{pmatrix}
	\begin{pmatrix}
	a_{b(i+1)+1} & a_{b(i+1)+2} & \cdots & a_{b(i+2)} \\
	a_{b(i+1)} & a_{b(i+1)+1} & \cdots & a_{b(i+2)-1} \\
	\vdots  & \vdots & \ddots & \vdots \\
	a_{bi+2} & a_{bi+3} & \cdots & a_{b(i+1)+1}
	\end{pmatrix}
	\end{displaymath}
	One can notice that the coefficients $a_{bi+2}, \cdots, a_{b(i+1)}$ \emph{overlap} between the two matrices: in particular, they occur respectively \emph{above} the main diagonal of $M_{F,i}$ and \emph{below} the main diagonal of $M_{F,i+1}$. As a consequence $L_{b,k,q}$ for $k>3$ is lower than $\left[ (q-1)q^{2(b-1)} \right]^{k-2}$, which is the number of ways one can choose a set of $k-2$ invertible Toeplitz matrices of size $b \times b$ over $\F_q$ with repetitions.
\end{remark}

We now model the problem of determining $L_{b,k,q}$ in terms of \emph{determinant} functions. Let $M_a$ be the Toeplitz matrix defined by the vector of $2b-1$ binary coefficients $a=(a_2,\cdots,a_{2b}) \in \F_q^{2b-1}$, and let us define $det:\F_q^{2b-1} \rightarrow \F_q$ as the function of $2b-1$ variables that associates to each vector $a \in \F_q^{2b-1}$ the determinant of the matrix $M_a$. Thus, we have that the support set $supp(det) = \{a \in \F_q^{2b-1}: det(a) \neq 0\}$ contains the vectors $a$ that define all $b\times b$ non-singular Toeplitz matrices. By Theorem~\ref{thm:num-lat-cub} it follows that the cardinality of the support is $|supp(det)| = q^{2(b-1)}(q-1)$.

Recall that by Remark~\ref{rem:overlap} the elements of two adjacent invertible Toeplitz matrices overlap respectively on the upper and lower triangular parts. Hence, in order to construct a Latin hypercube of dimension $k$, one can choose from the support of the determinant function $det$ a \emph{sequence} of $k-2$ vectors so that each each pair of adjacent vectors overlap respectively on the last and the first $b-1$ coordinates. We now formalize this reasoning in terms of \emph{de Bruijn graphs}. Let $s \in \N$ and $A$ be a finite alphabet, with $A^*$ denoting the free monoid of words over $A$. Given $u,v \in A^*$ such that $|u| \ge s$ and $|v| \ge s$, we define the $s$\emph{-fusion operator} $\odot$ as in~\cite{sutner}, that is $u \odot v = z$ if and only if there exists $x \in A^s$ and $u_0,v_0 \in A^*$ such that $u = u_0x$,  $v = xv_0$, and $z = u_0xv_0$. In other words $z$ is obtained by \emph{overlapping} the right part of $u$ and the left part of $v$ of length $s$. Setting $A = \F_q$ and $s = b-1$, we obtain the $s$-fusion operator for our case of overlapping vectors $u,v \in supp(det)$ of
length $2b-1$. The set of overlapping relations can be conveniently described using the de Bruijn graph associated to the determinant function, which we formally define below:
\begin{definition}
	\label{def:db-graph}
	Let $det: \F_q^{2b-1} \rightarrow \F_q$ be the determinant function for $b\times b$ Toeplitz matrices over $\F_q$. The \emph{de Bruijn graph} associated to $det$ is the directed graph $G_{det} = (V,E)$ where the set of vertices is $V = supp(det)$, while an ordered pair of vertices $(v_1,v_2)$ belongs to the set of edges $E$ if and only if there exists $z \in \F_2^{3b-1}$ such that $z = v_1 \odot v_2$, where $\odot$ denotes the $s$-fusion operator with $s = b-1$.
\end{definition}
\begin{example}
	\label{ex:det-tt}
	Let $b=2$, $k=5$ and $q=2$. In this case, the local rule  $f: \F_2^9 \rightarrow \F_2$ has coefficients $a_1$ and $a_9$ set to $1$, while the central coefficients $(a_{2i}, a_{2i+1}, a_{2i+2})$ define the following Toeplitz matrix for $i \in \{1,2,3\}$:
	\begin{equation}
	\label{eq:toep-mat-ex}
	M_{F},i = 
	\begin{pmatrix}
	a_{2i+1} & a_{2i+2} \\
	a_{2i} & a_{2i+1}
	\end{pmatrix}
	\end{equation}
	In particular $M_{F,1}$, $M_{F,2}$ and $M_{F,3}$ will be the matrices respectively associated to the second, third and fourth block. It is easily seen that the determinant of $M_{F,i}$ is $a_{2i+1} \oplus a_{2i}a_{2i+2}$. Figure~\ref{fig:tt-db-ex} reports the truth table and the de Bruijn graph of the determinant function $det: \F_2^3 \rightarrow \F_2$. Rows in bold in the table correspond to the vectors of the support of $det$, which in turn are the vertices of the de Bruijn graph.
	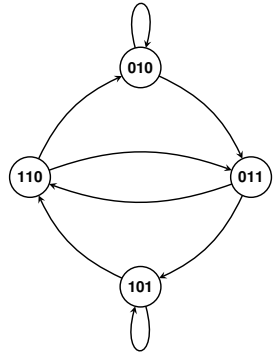
\begin{figure}[t]
		\centering
		\begin{subfigure}{.5\textwidth}
			\centering
			\begin{tabular}{cccc}
				\hline
				$a_{2i}$ & $a_{2i+1}$ & $a_{2i+2}$ & $a_{2i+1} \oplus a_{2i}a_{2i+2}$ \\
				\hline
				0 & 0 & 0    & 0 \\
				1 & 0 & 0    & 0 \\
				{\bfseries 0} & {\bfseries 1} & {\bfseries 0} & {\bfseries 1} \\
				{\bfseries 1} & {\bfseries 1} & {\bfseries 0} & {\bfseries 1} \\
				0 & 0 & 1    & 0 \\
				{\bfseries 1} & {\bfseries 0} & {\bfseries 1} & {\bfseries 1} \\
				{\bfseries 0} & {\bfseries 1} & {\bfseries 1} & {\bfseries 1} \\
				1 & 1 & 1    & 0 \\
				\hline
			\end{tabular}
		\end{subfigure}%
		\begin{subfigure}{.5\textwidth}
			\footnotesize
			\centering
			\resizebox{!}{5cm}{%
				\begin{tikzpicture}
				[->,auto,node distance=1.5cm, every loop/.style={min distance=12mm},
				empt node/.style={font=\bfseries,inner sep=0pt,outer sep=0pt},
				circ node/.style={circle,thick,draw,font=\sffamily\bfseries,minimum
					width=0.8cm, inner sep=0pt, outer sep=0pt}]
				
				\node [empt node] (e1) {};
				\node [circ node] (n00) [above=1.75cm of e1] {010};
				\node [circ node] (n01) [right=1.75cm of e1] {011};
				\node [circ node] (n10) [left=1.75cm of e1] {110};
				\node [circ node] (n11) [below=1.75cm of e1] {101};       
				
				\draw [->, thick, shorten >=0pt,shorten <=0pt,>=stealth] (n00) 
				edge[bend left=20] node (f5) [above right]{} (n01);
				\draw [->, thick, shorten >=0pt,shorten <=0pt,>=stealth] (n01)
				edge[bend left=20] node (f5) [below right]{} (n11);
				\draw [->, thick, shorten >=0pt,shorten <=0pt,>=stealth] (n11)
				edge[bend left=20] node (f5) [below left]{} (n10);
				\draw [->, thick, shorten >=0pt,shorten <=0pt,>=stealth] (n10)
				edge[bend left=20] node (f5) [above left]{} (n00);
				\draw[->, thick, shorten >=0pt,shorten <=0pt,>=stealth] (n10) edge[bend
				left=20] node (f1) [above]{} (n01);
				\draw[->, thick, shorten >=0pt,shorten <=0pt,>=stealth] (n01)
				edge[bend left=20] node (f2) [below]{} (n10);
				\draw[->, thick, shorten >=0pt,shorten <=0pt,>=stealth] (n00) edge[loop
				above] node (f3) [above]{} ();
				\draw[->, thick, shorten >=0pt,shorten <=0pt,>=stealth] (n11) edge[loop
				below] node (f4) [below]{} ();
				\end{tikzpicture}
			}
		\end{subfigure}
		\caption{Truth table (left) and de Bruijn graph (right) of $a_{2i+1} \oplus a_{2i}a_{2i+2}$.}
		\label{fig:tt-db-ex}
	\end{figure}
	Following Figure~\ref{fig:tt-db-ex}, one can construct the local rule $f$ by choosing a sequence of $k-2=3$ vectors from the support of $det(a_2,a_3,a_4)$ such that both the first and the second and the second and the third overlap respectively on the last and the first bit. This actually amounts to finding a path of length $2$ on the de Bruijn graph. An example could be the sequence $(0,1,0) - (0,1,1) - (1,0,1)$. In particular, $(0,1,0)=(a_2,a_3,a_4)$ is the vector defining the Toeplitz matrix of the second block. Similarly, the vector $(0,1,1)=(a_4,a_5,a_6)$ defines the Toeplitz matrix of the third block while $(1,0,1)=(a_6,a_7,a_8)$ defines the Toeplitz matrix of the fourth block. Consequently, the local rule $f:\F_2^9 \rightarrow \F_2$ is defined as $f(x_1,\cdots,x_9) = x_1 \oplus x_3 \oplus x_5 \oplus x_6 \oplus x_8 \oplus x_9$, and by Theorem~\ref{thm:lat-hc-ca} the corresponding LBCA $F:\F_2^{10} \rightarrow \F_2^2$ generates a $5$-dimensional Latin hypercube of order $4$.
\end{example}

\section{Counting Sequences of Invertible Matrices}
\label{sec:count}
As we said above, the de Bruijn graph of the determinant function summarizes all the overlap relations between vectors of its support, which represent invertible Toeplitz matrices. Thus, we derived that Latin hypercubes of dimension $k$ generated by LBCA correspond to \emph{paths of length} $k-3$ over this graph:
\begin{theorem}
	\label{thm:count-hc}
	Let $b,k \in \N$. Then, $L_{b,k}$ equals the number of paths of length $k-3$ over the de Bruijn graph $G_{det}$ of the determinant function $det:\F_q^{2b-1} \rightarrow \F_q$.
\end{theorem}
Hence, in order to count the number of $k$-dimensional Latin hypercubes generated by LBCA, we need to look more closely at the properties of the de Bruijn graph $G_{det}$ associated to the determinant of Toeplitz matrices. In particular, counting the number of paths of length $k-3$ over $G_{det}$ requires characterizing the indegrees and outdegrees of its vertices. Looking at Figure~\ref{fig:tt-db-ex} one can see that each vertex has two ingoing and two outgoing edges, hence the resulting de Bruijn graph $G_{det}$ for $b=2$ and $q=2$ is $2$-\emph{regular}. In the remainder of this section we will show that a regularity property holds in general for every $b\ge2$ and $q$ power of a prime. We first need the following result:
\begin{theorem}
	\label{thm:toep-restr}
	Denote by $T(n)$ the set of $n\times n$ Toeplitz matrices over $\F_q$. Let $A \in T(n)$ be strictly lower triangular, then there are exactly $(q - 1)q^{n-1}$ upper triangular matrices $B \in T(n)$ such that $A + B$ is nonsingular.
	\begin{proof}
		The result is clear if $A$ is the all-zero matrix, therefore we assume henceforth that A is nonzero. We shall use the results by Daykin~\cite{daykin} on \emph{persymmetric} (Hankel) matrices. Clearly, results on persymmetric matrices can be restated in terms of Toeplitz matrices, since the former are the transpose of the latter. Let $H(b)$ be the set of $n \times n$ persymmetric matrices. For any $A \in H(b)$ with $a_{i,j} = 0$ if $i + j \ge b + 1$, denote the number of matrices $B \in H(b)$ with $b_{i,j} = 0$ if $i + j \le n$ such that $A + B$ is nonsingular as $N(A)$. We thus show that $N(A) = (q-1)q^{n-1}$.
		
		For any $M \in H(n)$ and any $m \le n$, we denote the matrix in $H(m)$ consisting of the first $m$ rows and $m$ columns of $M$ as $M[m]$. Let $P \in H(m)$ be a nonzero matrix. Let $R(P)$ be the number of nonsingular matrices $Q \in H(2m)$ such that $P = Q[m]$, and for all $m$ and $i \le m$, let $T(m : i) = (q-1)^2 q^{2m-i-2}$ if $i<m$ and $T(m : i) = (q-1) q^{m-1}$ if $i= m$. Theorem 3 in~\cite{daykin} gives the following formulas for $R(P)$ and $R'(P)$:
		\begin{displaymath}
		R(P) = 
		\begin{cases}
		\frac{T(2m-v : m-v)}{(q-1)q^{m-v-1}} & \mbox{if } v<m \\
		q \sum_{i=1}^m T(m : i) & \mbox{otherwise } 
		\end{cases}
		, \enspace R'(P) = 
		\begin{cases}
		\frac{T(2m-v : m-v)}{(q-1)q^{m-v-1}} & \mbox{if } v<m \\
		q \sum_{i=1}^m T(m : i) & \mbox{otherwise } 
		\end{cases}
		\end{displaymath}
		We now show that for all nonzero $P \in H(m)$, $R(P) = (q-1)q^{2m-1}$. We prove the claim for $R(P)$. If $v < m$, we have
		\begin{displaymath}
		R(P) = \frac{T(2m-v : m-v)}{(q-1)q^{m-v-1}} = \frac{(q-1)^2 q^{2(2m-v)-(m-v)-2}}{(q-1)q^{m-v-1}} = (q-1)q^{2m-1} \enspace .
		\end{displaymath}
		If $v=m$ we have
		\begin{align*}
		R(P) &= q \sum_{i=1}^m T(m : i) = q(q-1) \left\{ (q-1)q^{m-1} \sum_{i=1}^{m-1} q^{m-1-i} + q^{m-1}\right\} = \\ &= q^m(q-1) \{ (q^{m-1} - 1) + 1\} = (q-1)q^{2m-1} \enspace .
		\end{align*}
		With a similar argument, one can also show that $R'(P) = (q-1)q^{2m-1}$. Now, let $m = \lceil n/2 \rceil$ and $P = A[m]$. If $n = 2m$ is even, then $N(A) = R(P) = (q - 1)q^{n-1}$. If $n = 2m-1$ is odd then $P_{m,m} = 0$, thus consider the matrices $P^a$, obtained by setting the value $P^a_{m,m} = a$, for all $a \in \F_q$. Then $N(A) = \sum_{a \in \F_q} R(P^a) = (q-1)q^{n-1}$. \qed
	\end{proof}
\end{theorem}
Theorem~\ref{thm:toep-restr} thus states that by fixing the leftmost $b-1$ entries of the vector $(a_2,\cdots,a_{2b})$ that defines a Toeplitz matrix $A$, one can complete the remaining $b$ ones in $(q-1)q^{b-1}$ different ways so that the resulting Toeplitz matrix is invertible. This brings us to the following corollary:
\begin{corollary}
	\label{cor:bal-restr}
	Let $det:\F_q^{2b-1}\rightarrow \F_q$ be the determinant function associated to the set $T(b)$. Then, for any vector $\tilde{a} \in \F_q^{b-1}$, the restriction $det|_{\tilde{a}}:\F_q^b \rightarrow \F_q$ obtained by fixing
	either the leftmost or the rightmost $b-1$ coordinates to $\tilde{a}$ is balanced, that is $|supp(det|_{\tilde{a}})| = (q-1)q^{b-1}$.
\end{corollary}
\noindent
We now show that this corollary implies the regularity of $G_{det}$.
\begin{lemma}
	\label{lm:reg-dbg}
	For any $b \ge 2$ the de Bruijn graph $G_{det}$ of the determinant function $det$ is $(q-1)q^{b-1}$-regular.
	\begin{proof}
		As a preliminary remark, observe that by Theorem~\ref{thm:num-lat-cub} the number of vertices in $G_{det}$ is $|V|=(q-1)q^{2(b-1)}$. We prove only that the outdegree of each vertex is $(q-1)q^{b-1}$, the indegree case following from a symmetrical reasoning. Let us fix the first $b-1$ coordinates $a_2,\cdots,a_{b}$ of $det$ to a vector $\tilde{a} \in \F_2^{b-1}$. Since the restriction $det|_{\tilde{a}}$ of $b$ variables induced by $\tilde{a}$ is balanced, it means that there is a set $_{\tilde{a}}V = \{v \in V: (v_1, \cdots, v_{b-1}) = \tilde{a} \}$ of $(q-1)q^{b-1}$ vertices in $G_{det}$ that begins by $\tilde{a}$. Each vertex in $G_{det}$ that ends by $\tilde{a}$ has an outgoing degree of $q(q-1)^{b-1}$, since it overlaps with all vertices in $_{\tilde{a}}V$. Let $V_{\tilde{a}} = \{v \in V: (v_{b+1}, \cdots, v_{2b-1}) = \tilde{a} \}$ be the set of all such vertices. By Corollary~\ref{cor:bal-restr} the restriction $det|_{\tilde{a}}$ obtained by fixing the \emph{last} $b-1$ coordinates to $\tilde{a}$ is also balanced. Thus, the cardinality of $V_{\tilde{a}}$ is also $(q-1)q^{b-1}$, meaning that there are $(q-1)q^{b-1}$ vertices ending by $\tilde{a}$ that have outdegree $(q-1)q^{b-1}$. Since this property holds for any vector $\tilde{a} \in \F_q^{b-1}$, it follows that there are $q^{b-1}\cdot(q-1)q^{b-1} = (q-1)q^{2(b-1)} = |V|$ distinct vertices with outdegree $(q-1)q^{b-1}$. \qed
	\end{proof}
\end{lemma}
Using Lemma~\ref{lm:reg-dbg}, we can now determine what is the number of $k$-dimensional Latin hypercubes of order $q^b$ generated by LBCA:
\begin{theorem}
	\label{thm:num-hc-ca}
	Let $b, k \in \N$, with $k\ge 3$. Then, the number of $k$-dimensional Latin hypercubes of order $q^b$ generated by LBCA $F: \F_q^{bk} \rightarrow \F_q^b$ with local rules $f: \F_{q}^{b(k-1)+1}\rightarrow \F_q$ is $L_{b,k,q} = (q-1)^{k-2}q^{(k-1)(b-1)}$.
	\begin{proof}
		By Theorem~\ref{thm:count-hc} $L_{b,k,q}$ equals the number of paths of length $k-3$ over the de Bruijn graph $G_{det}$. We shall prove the result by induction on $k$.
		
		For $k=3$, the number of paths of length $0$ over $G_{det}$ obviously coincides with the number of vertices, which is $(q-1)q^{2(b-1)}$ by Corollary~\ref{thm:num-lat-cub}.
		
		Assume now that $k>3$, and let us consider $L_{b,k+1, q}$. Clearly, the paths of length $k+1$ are constructed by adding a new edge to all paths of length $k$, which by induction hypothesis are $(q-1)^{k-2}q^{(k-1)(b-1)}$. Since $G_{det}$ is $(q-1)q^{b-1}$-regular, we thus have
		\begin{displaymath}
		L_{b,k+1,q} = (q-1)^{k-2}q^{(k-1)(b-1)}\cdot (q-1)q^{(b-1)} = (q-1)^{k-1}q^{k(b-1)} \enspace .
		\end{displaymath}\qed
	\end{proof}
\end{theorem}

\section{Conclusions}
\label{sec:conc}
In this paper, we addressed the construction of Latin hypercubes generated by LBCA over the finite field $\F_q$, thereby taking a first step towards the generalization of the results in~\cite{mariot20} about CA-based mutually orthogonal Latin squares. More precisely, we generalized the block construction of~\cite{mariot20} to dimension $k>2$, showing that the permutation property between any of the central $k-2$ blocks of the CA and the final configuration is related to the invertibility of the Toeplitz matrices defined by the central coefficients of the local rule. Moreover, we observed that the Toeplitz matrices associated to adjacent blocks share the coefficients respectively on the upper and lower triangulars, a property that can be described by the de Bruijn graph of the determinant function. We finally derived the number $L_{b,k,q}$ of LBCA generating $k$-dimensional Latin hypercubes of order $q^b$ by counting the number of paths of length $k-3$ over this de Bruijn graph, which we proved to be $(q-1)q^{b-1}$-regular. The resulting formula shows that $L_{b,k,q}$ is exponential both in the dimension and the block size of the hypercube, therefore indicating that the family of Latin hypercubes generated by LBCA is quite large. We plan to investigate this rich structure of Latin hypercubes in future research, in particular by characterizing its mutually orthogonal subsets.


\bibliographystyle{abbrv}
\bibliography{bibliography}

\begin{thebibliography}{10}

\bibitem{blakley79}
G.~R. Blakley.
\newblock Safeguarding cryptographic keys.
\newblock In {\em Managing Requirements Knowledge, International Workshop on},
  pages 313--317, 1979.

\bibitem{daykin}
D.~Daykin.
\newblock Distribution of bordered persymmetric matrices in a finite field.
\newblock {\em J. Reine Angew. Math.(Crelle’s J.)}, 203:47--54, 1960.

\bibitem{armas11}
M.~Garc{\'\i}a-Armas, S.~R. Ghorpade, and S.~Ram.
\newblock Relatively prime polynomials and nonsingular {H}ankel matrices over
  finite fields.
\newblock {\em Journal of Combinatorial Theory, Series A}, 118(3):819--828,
  2011.

\bibitem{herranz18}
J.~Herranz and G.~S{\'{a}}ez.
\newblock Secret sharing schemes for (k, n)-consecutive access structures.
\newblock In {\em {CANS}}, volume 11124 of {\em LNCS}, pages 463--480.
  Springer, 2018.

\bibitem{mariot20}
L.~Mariot, M.~Gadouleau, E.~Formenti, and A.~Leporati.
\newblock Mutually orthogonal latin squares based on cellular automata.
\newblock {\em Des. Codes Cryptogr.}, 88(2):391--411, 2020.

\bibitem{mariot14}
L.~Mariot and A.~Leporati.
\newblock Sharing secrets by computing preimages of bipermutive cellular
  automata.
\newblock In {\em Cellular Automata - 11th International Conference on Cellular
  Automata for Research and Industry, {ACRI} 2014, Krakow, Poland, September
  22-25, 2014. Proceedings}, pages 417--426, 2014.

\bibitem{mariot17}
L.~Mariot, A.~Leporati, A.~Dennunzio, and E.~Formenti.
\newblock Computing the periods of preimages in surjective cellular automata.
\newblock {\em Nat. Comput.}, 16(3):367--381, 2017.

\bibitem{mariot19}
L.~Mariot, S.~Picek, A.~Leporati, and D.~Jakobovic.
\newblock Cellular automata based s-boxes.
\newblock {\em Cryptography and Communications}, 11(1):41--62, 2019.

\bibitem{shamir79}
A.~Shamir.
\newblock How to share a secret.
\newblock {\em Commun. {ACM}}, 22(11):612--613, 1979.

\bibitem{stinson04}
D.~R. Stinson.
\newblock {\em Combinatorial designs - constructions and analysis}.
\newblock Springer, 2004.

\bibitem{sutner}
K.~Sutner.
\newblock De {B}ruijn graphs and linear cellular automata.
\newblock {\em Complex Systems}, 5(1):19--30, 1991.

\end{thebibliography}

\end{document}